# Application Specific Hardware Design Simulation for High Performance Embedded System

Ravi Khatwal
Research scholar
Department Of Computer science,
Mohan LaL Sukhadia University,
Udaipur, India.

Manoj Kumar Jain, Ph.D
Associate Professor,
Department Of Computer science,
Mohan LaL Sukhadia University,
Udaipur, India.

## ABSTRACT
Application specific simulation is challenging task in various real time high performance embedded devices. In this study specific application is implemented with the help of Xilinx. Xilinx provides SDK and XPS tools, XPS tools used for develop complete hardware platform and SDK provides software platform for application creation and verification. Xilinx XUP-5 board have been used and implemented various specific Applications with hardware platform. In this study the base instruction set with customized instructions, supported with specific hardware resources are analyzed.

## Keywords
Xilinx, virtex-5 FPGA board, simulation, hardware and software design, Xilinx Platform Studio.

## 1. INTRODUCTION
In co-design methodology, the hardware and software components for an embedded system are designed jointly. Each of the hardware and software components designed using appropriate tools (hardware synthesis, code generation and hardware-software co-simulation tools). In ASIP design technology hardware can be design according for specific application. Xilinx SDK provides software environment used for various specific application verification and creation. Kucukcakar, K. [1] proposed a unique architecture and methodology to design ASIPs in the embedded controller domain by customizing an existing processor instruction set and architecture. Jain, M.K., Balakrishnan, M., Kumar, A. [2] proposed survey in ASIP area and identifies some issues which need to be addressed. Hartmann, M., Raghavan, P., Agrawal,P., Dehaene, W. [3] proposed a design method for memristor-based (ReRAM) memory architectures for embedded processors to address the effects caused by longer write latencies. Sharma, A., Sutar, S., Sharma, V.K., Mahapatra, K.K. [4] designed an ASIP using language for instruction-set architecture (LISA) and designed processor has optimized instructions for the image enhancement application in spatial domain. Fathy, A.,Isshiki, T.,Li, D., Kunieda, H. [5] presented a complete framework for searching for Application specific special instruction patterns based on tree scan algorithm while tweaking it to fit real applications. J. Qiu, X. Gao, Y. Jiang, X. Xiao [7] proposed a hybrid simulation framework which improves the previous simulation methods by aggressively utilizing the host machine resources. H. M. Hassan, K. Mohammed and A. F. Shalash [8] presented an ASIP design for a discrete Fourier transform (DFT)/discrete cosine transform (DCT)/finite impulse response filters (FIR) engine.

## 2. EMBEDDED DEVELOPMENT KIT
Xilinx [6] provides Embedded Development Kit (EDK) (see figure 1) tools to design a complete embedded processor system for implementation in a Xilinx FPGA device. Xilinx Platform Studio (XPS) is the development environment used for designing the hardware platform and Software Development Kit (SDK) is an integrated development environment used for C/C++ embedded software application creation and verification. Embedded Development Kit also provides ISE Plan ahead is used to design custom Memory and processor configuration according to specific application. Xilinx EDK provides XPS, SDK and ISE tools for simulation analysis.

### 2.1 XPS
XPS provides hardware system development environments and provides Specification of the microprocessor, peripherals, and the interconnected components, along with their respective detailed configuration.

### 2.2 SDK
SDK provides software development environments and also used for developing standalone application.

### 2.3 ISE Design
ISE design used for verifies the correct functionality of Hardware Description Language (HDL), RTL and schematic design.

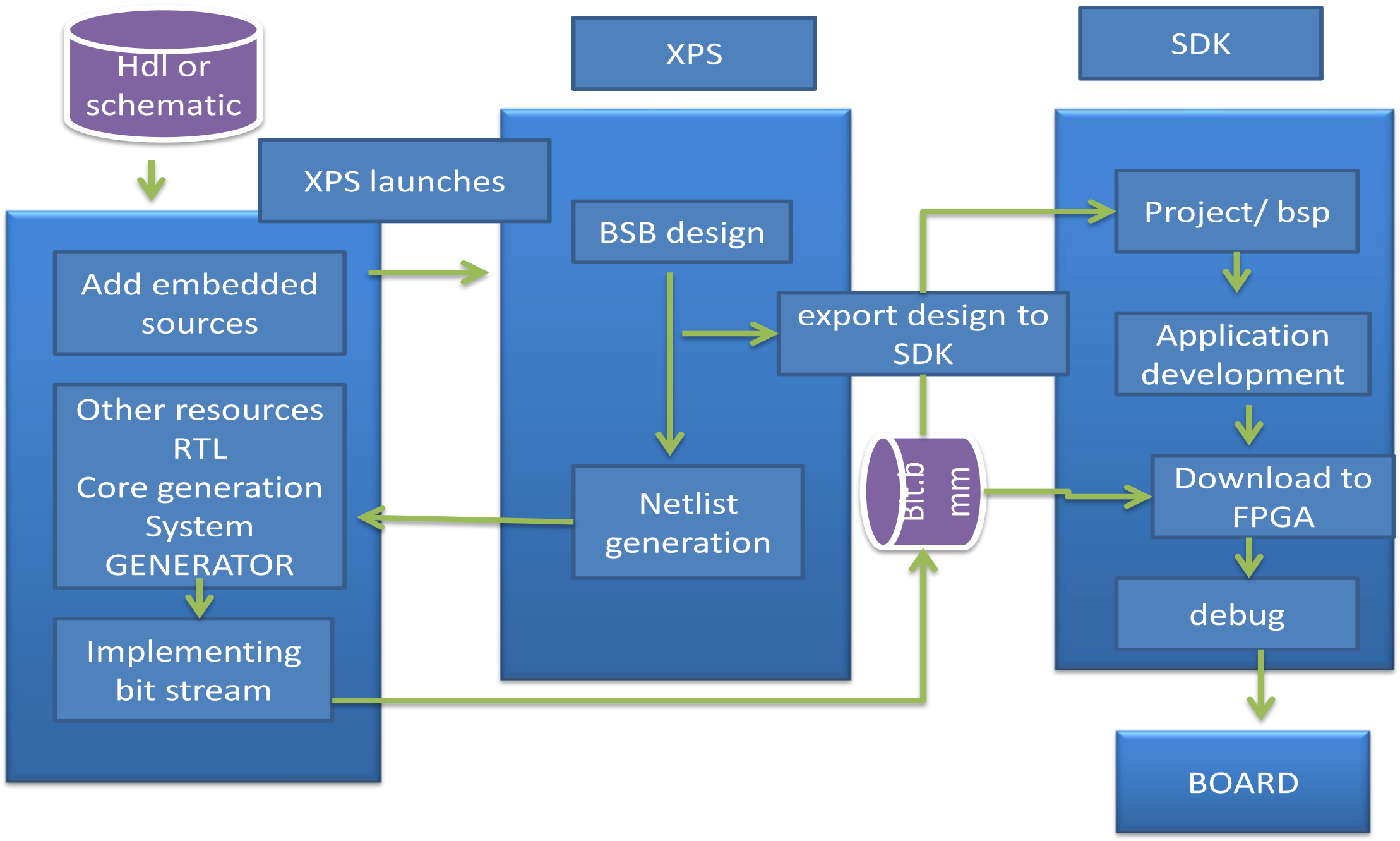

**Fig 1: EDK Design flow analysis**

## 3. XILINX PLATFORM STUDIO (XPS)

Xilinx Platform Studio (XPS) provides an interactive development environment that allows specifying all aspects of our hardware platform and XPS also maintains the hardware platform description in a high-level form, known as the Microprocessor Hardware Specification (MHS) file. XPS used to synthesizes the MHS source file into netlists used for the FPGA place and route process. The MHS file [6] is integral to our design process and contains all peripheral instantiations along with their parameters. The MHS file also defines the configuration of the embedded processor system and includes information on the bus architecture; peripherals, processor, connectivity, and address space etc. One of the supported embedded processor development boards available from Xilinx [6] has been selected for target architecture.

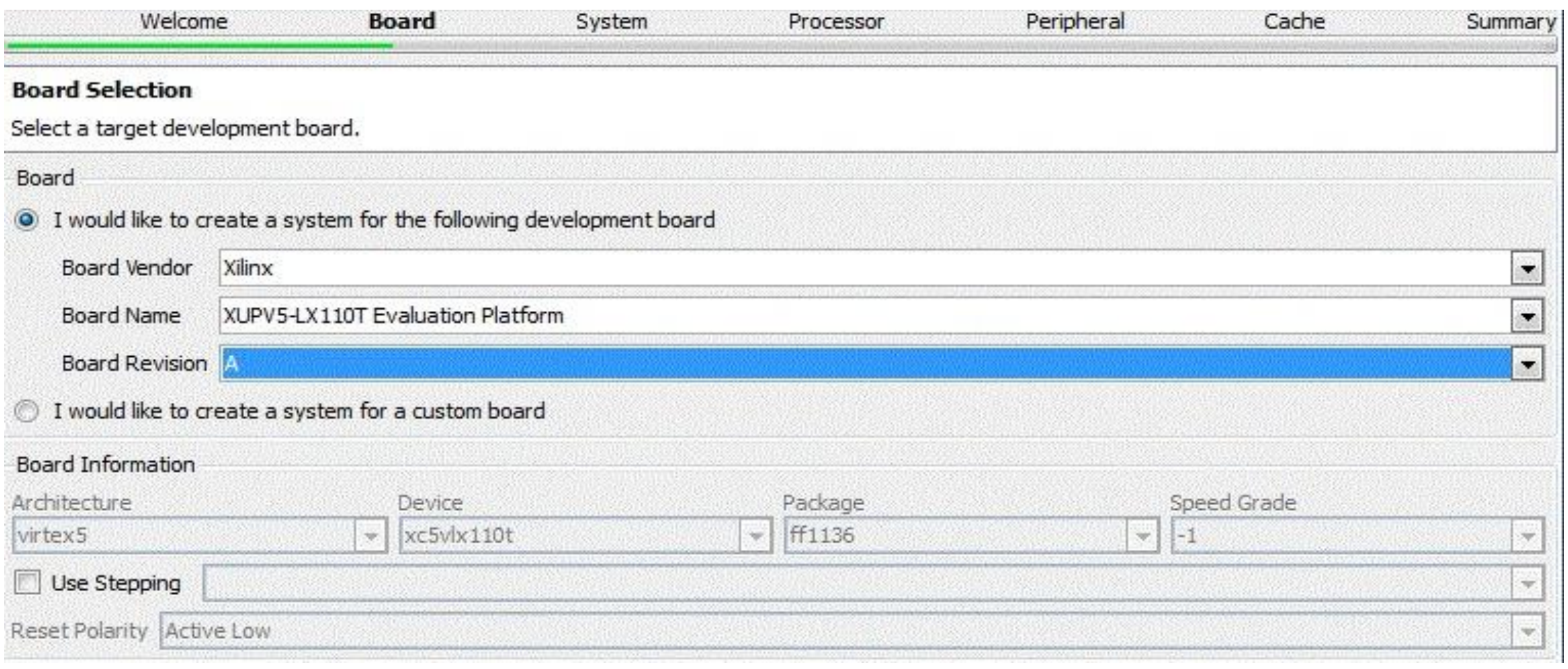

**Fig 2: xup-5 fpga board selection**

In XPS platform have configured selected XUP-5 FPGA platform (see figure2) for simulation and XUP-5-LX110T board utilizes Xilinx virtex5 XC5VLX110T-FF1136 device. A complete embedded processor system implemented within a Xilinx FPGA device. Hardware platform contains one or more processors, along with a variety of peripherals and memory blocks. The behavior of each processor or peripheral core can be easily customized (see figure 3,4&5). Hardware platform is ultimately implemented in the FPGA and Instruction and Data cache sized change according to desire application (see figure 6&7).

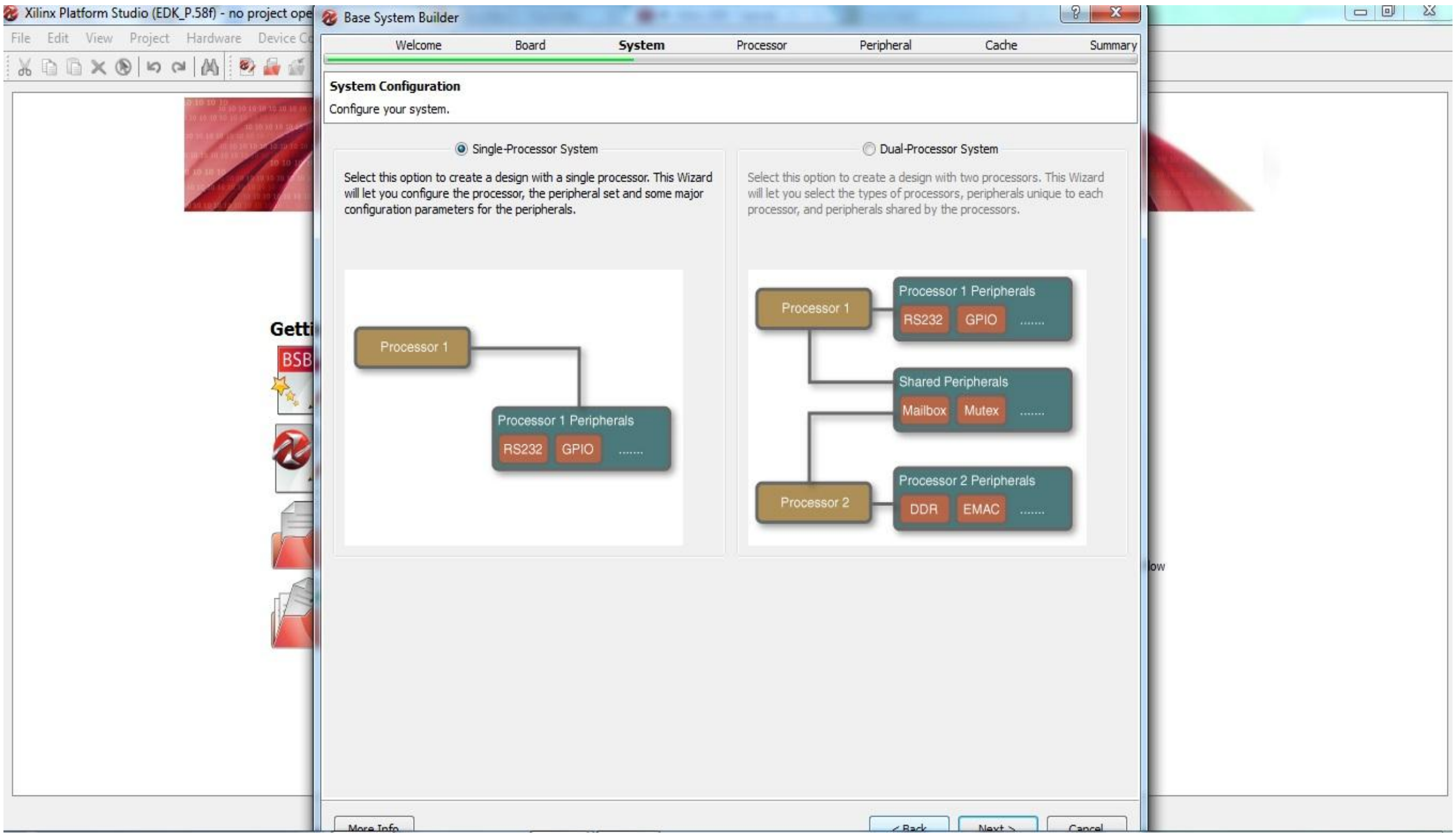

Fig 3: Processor configuration

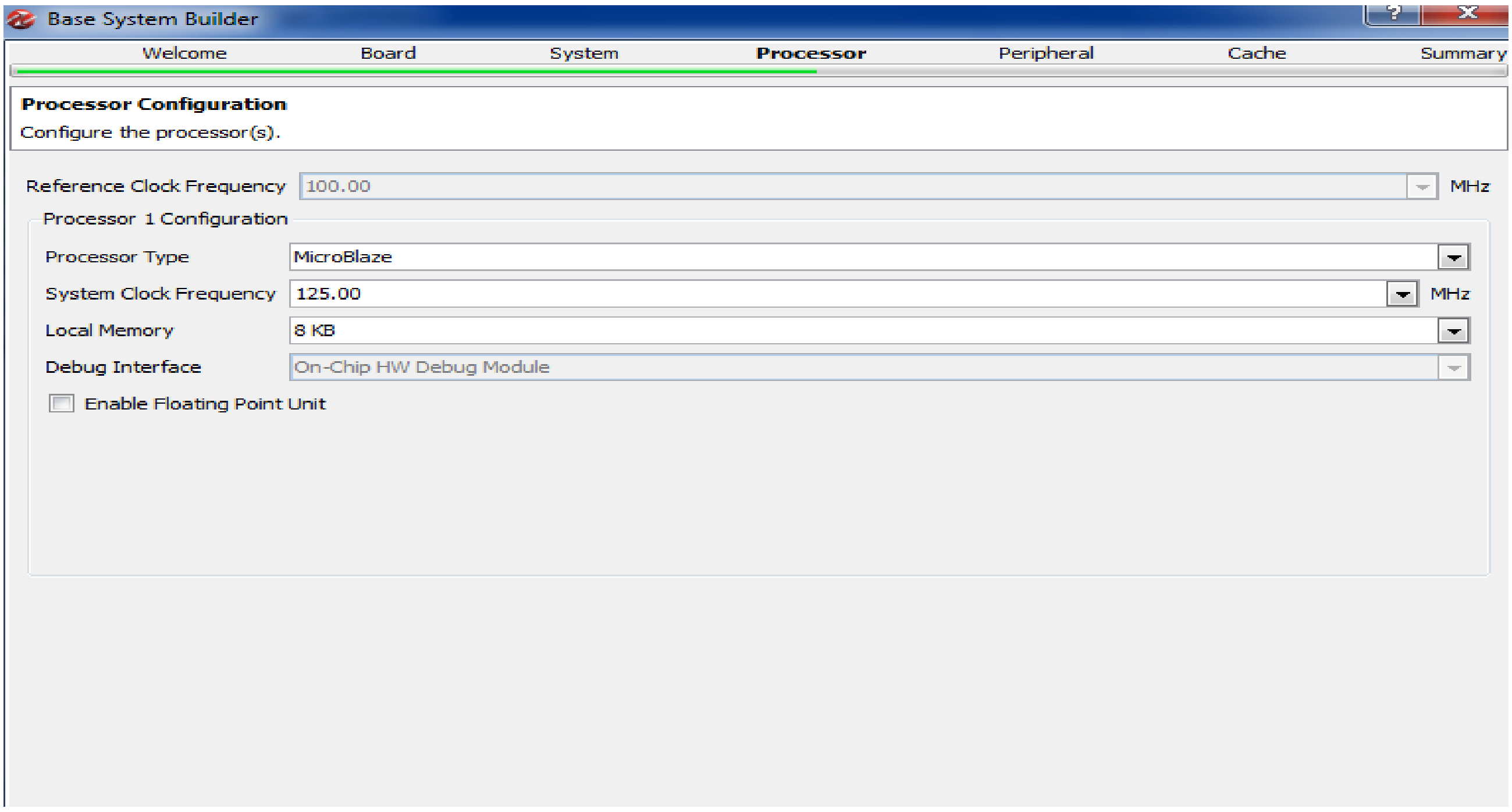

Fig 4: Micro blaze processor configuration

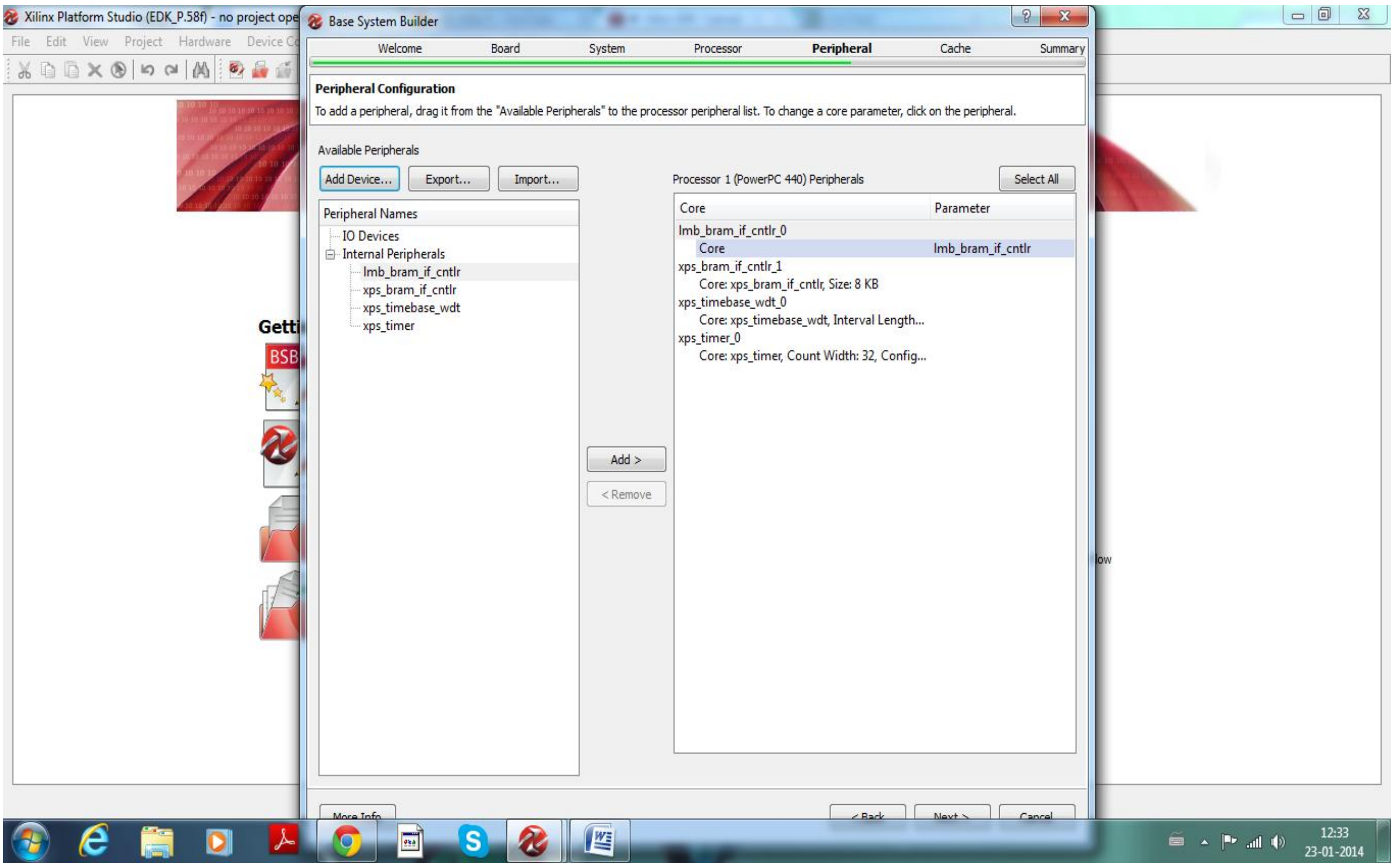

**Fig 5: Peripheral design**

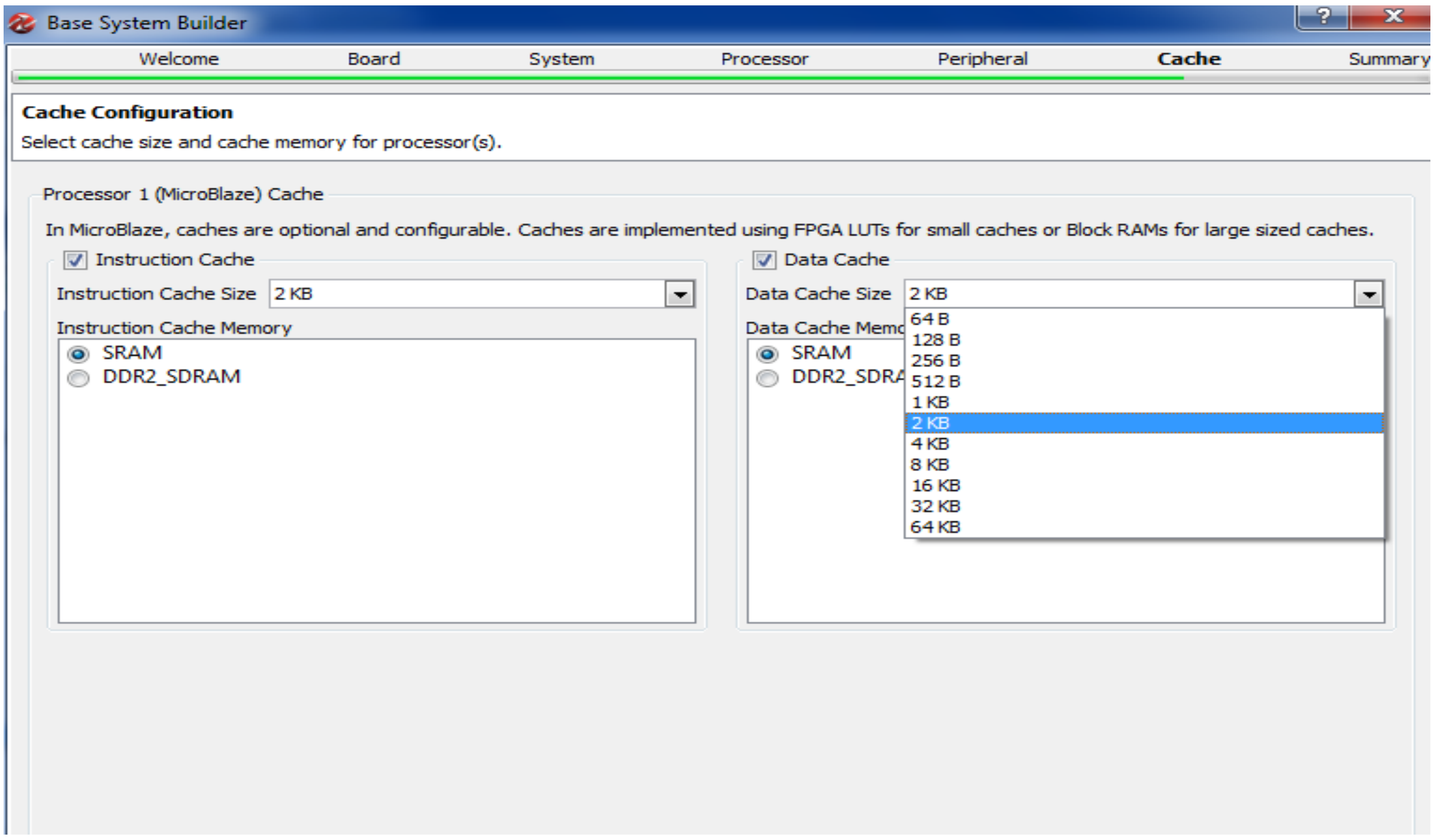

**Fig 6: I- Cache configuration**

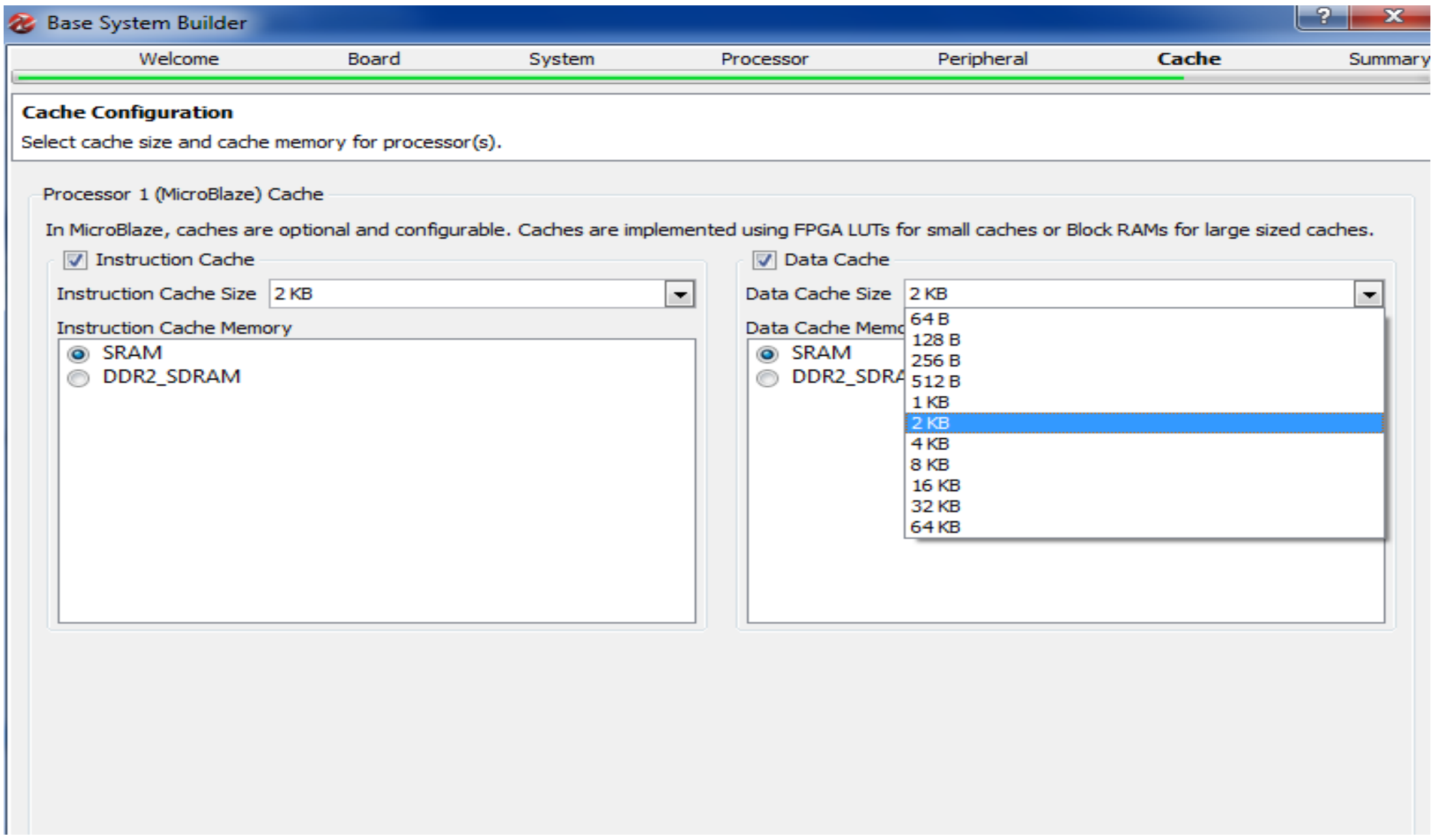

**Fig 7: D-Cache configuration**

XPS (see figure 8) used the Base System Builder, and it generated a bit stream for the FPGA. Hardware platform is exported to the Software Development Kit (SDK) platform. The hardware platform contains the XML-based hardware description file, the bit stream file, and the BMM file. XPS System Assembly View allows for view and configures the system block elements. XPS under implement flow; it generated net list implements, the design using the Xilinx backend, synthesis and route to create the final Netlist. The architectural design of selected processor and memory (see figure 9 & 11) and their VHDL configuration used for specific application (see figure 10 & 12).

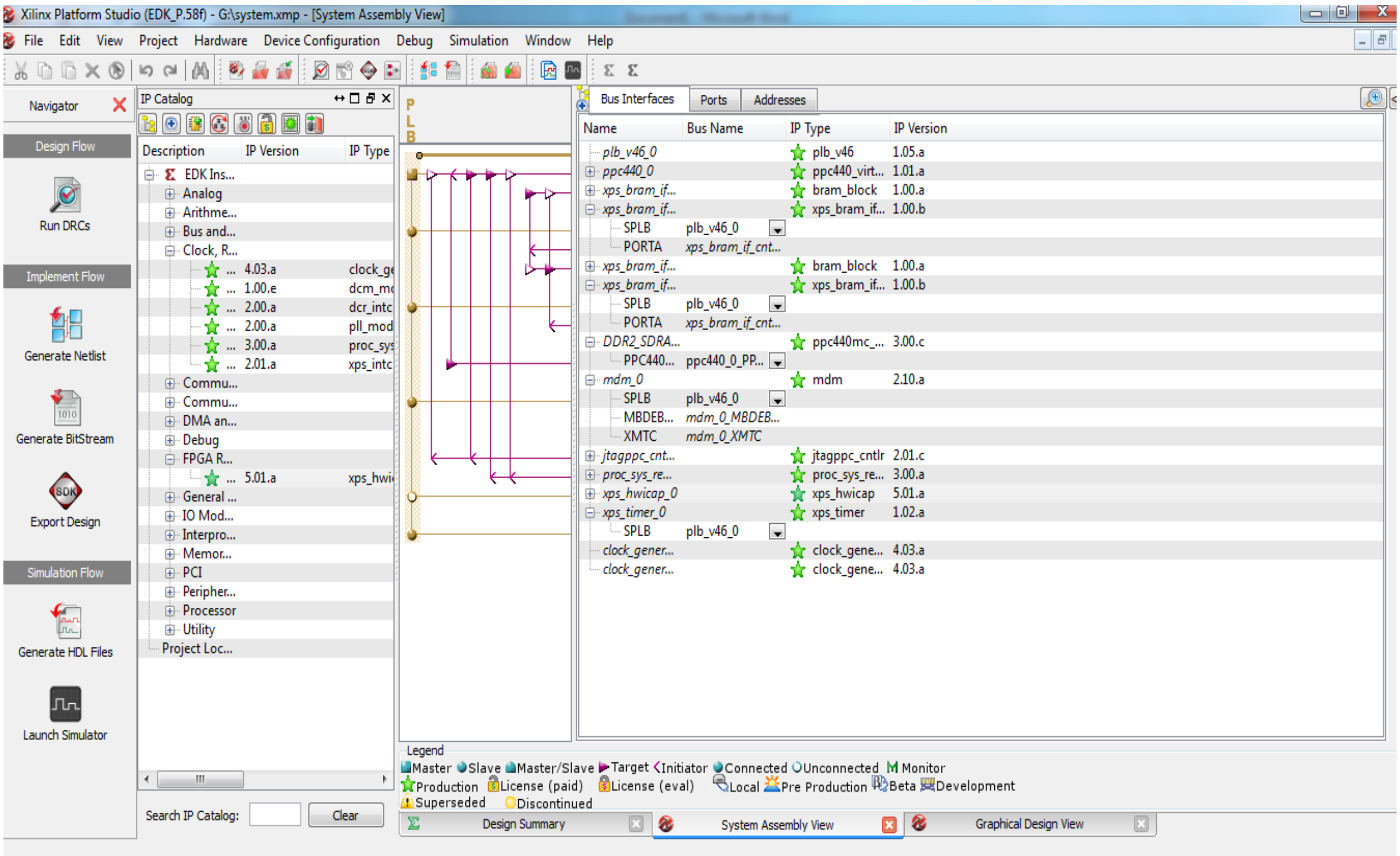

**Fig 8: XPS system assembly view analysis**

```
library UNISIM;
use UNISIM.VCOMPONENTS.ALL;

library xps_mch_emc_v3_01_a;
use xps_mch_emc_v3_01_a.all;

entity system_sram_wrapper is
  port (
    MCH_SPLB_Clk : in std_logic;
    RdClk : in std_logic;
    MCH_SPLB_Rst : in std_logic;
    MCH0_Access_Control : in std_logic;
    MCH0_Access_Data : in std_logic_vector(0 to 31);
    MCH0_Access_Write : in std_logic;
    MCH0_Access_Full : out std_logic;
    MCH0_ReadData_Control : out std_logic;
    MCH0_ReadData_Data : out std_logic_vector(0 to 31);
    MCH0_ReadData_Read : in std_logic;
    MCH0_ReadData_Exists : out std_logic;
    MCH1_Access_Control : in std_logic;
    MCH1_Access_Data : in std_logic_vector(0 to 31);
    MCH1_Access_Write : in std_logic;
    MCH1_Access_Full : out std_logic;
    MCH1_ReadData_Control : out std_logic;
    MCH1_ReadData_Data : out std_logic_vector(0 to 31);
    MCH1_ReadData_Read : in std_logic;
    MCH1_ReadData_Exists : out std_logic;
    MCH2_Access_Control : in std_logic;
    MCH2_Access_Data : in std_logic_vector(0 to 31);
    MCH2_Access_Write : in std_logic;
    MCH2_Access_Full : out std_logic;
    MCH2_ReadData_Control : out std_logic;
    MCH2_ReadData_Data : out std_logic_vector(0 to 31);
    MCH2_ReadData_Read : in std_logic;
    MCH2_ReadData_Exists : out std_logic;
    MCH3_Access_Control : in std_logic;
```

**Fig 9: Memory VHDL configuration files**

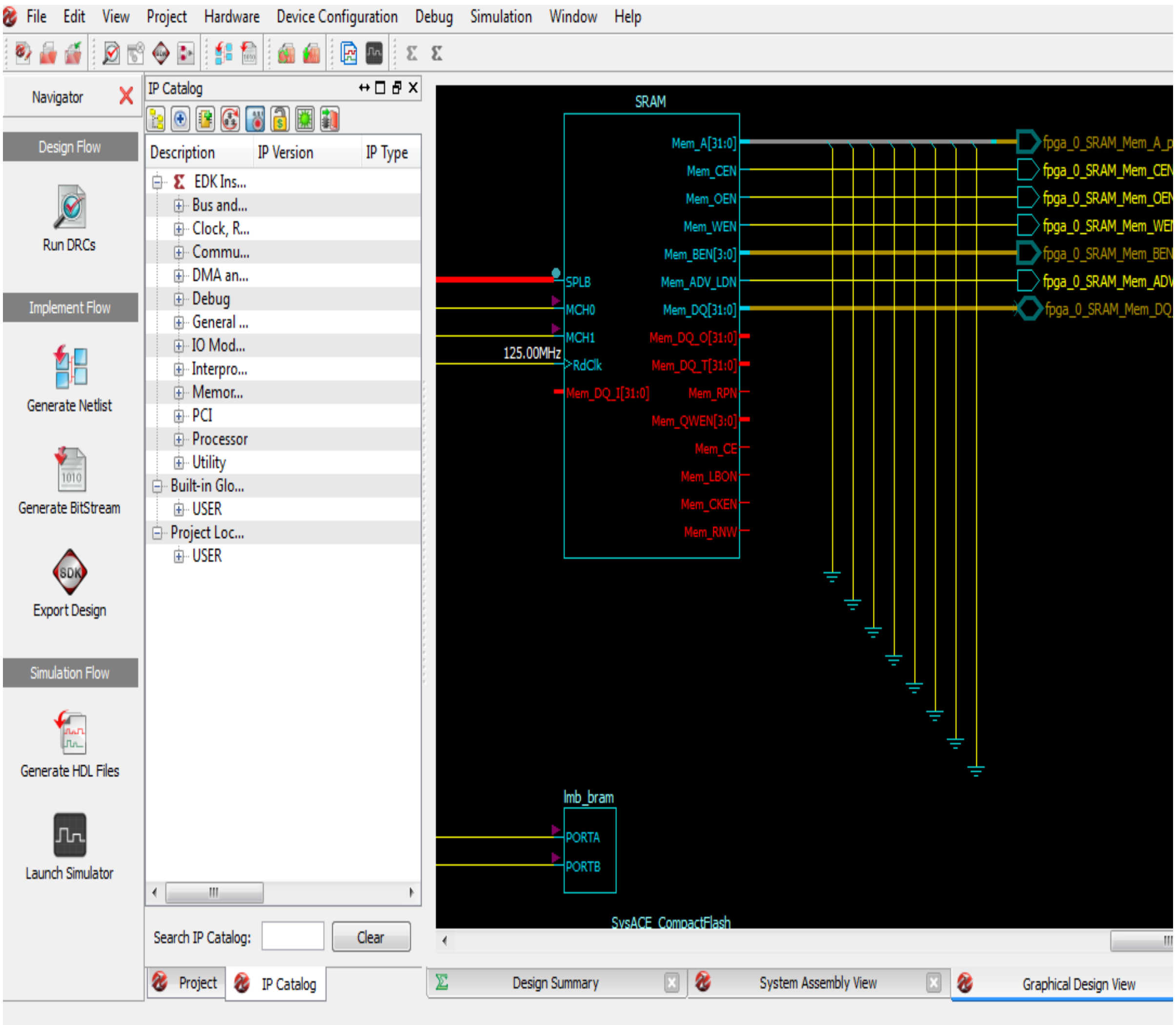

**Fig 10: Graphical Design of Memory**

```
architecture STRUCTURE of system_microblaze_0_wrapper is

  component microblaze is
    generic (
      C_SCO : integer;
      C_FREQ : integer;
      C_DATA_SIZE : integer;
      C_DYNAMIC_BUS_SIZING : integer;
      C_FAMILY : string;
      C_INSTANCE : string;
      C_AVOID_PRIMITIVES : integer;
      C_FAULT_TOLERANT : integer;
      C_ECC_USE_CE_EXCEPTION : integer;
      C_LOCKSTEP_SLAVE : integer;
      C_ENDIANNESS : integer;
      C_AREA_OPTIMIZED : integer;
      C_OPTIMIZATION : integer;
      C_INTERCONNECT : integer;
      C_STREAM_INTERCONNECT : integer;
      C_BASE_VECTORS : std_logic_vector;
      C_DPLB_DWIDTH : integer;
      C_DPLB_NATIVE_DWIDTH : integer;
      C_DPLB_BURST_EN : integer;
      C_DPLB_P2P : integer;
      C_IPLB_DWIDTH : integer;
```

**Fig 11: Processor configuration VHDL file**

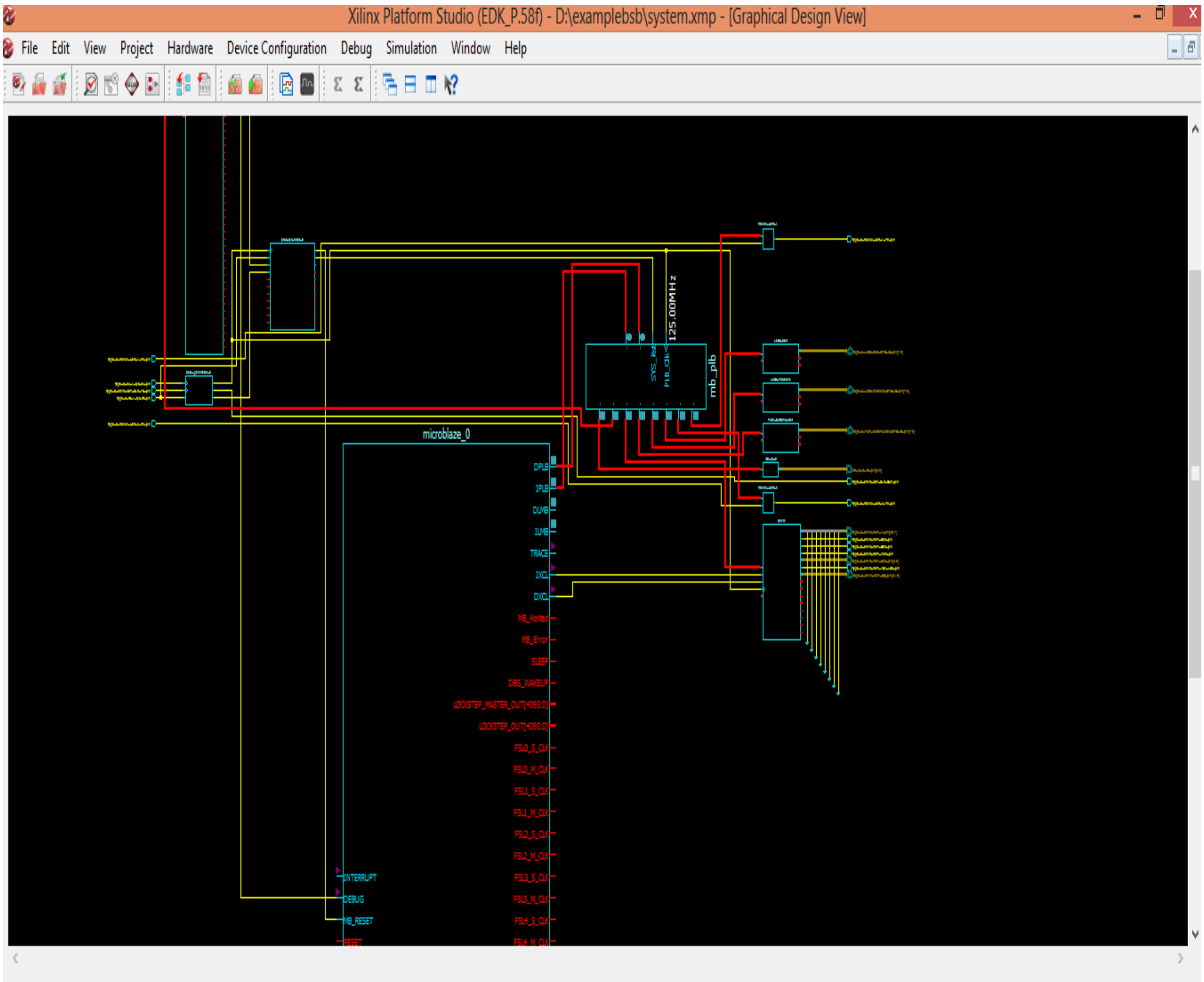

**Fig 12: Microblaze processor design**

## 4. SOFTWARE DEVELOPMENT KIT

XPS design is exported to the Software Development Kit (SDK) platform. Various Software applications must link against or run on top of a given software platform, using the specific provided Application Program Interfaces (APIs). Software Part (sdk) contains one or more source files, along with the necessary header files, used for compilation and generation of a binary output (.elf) file (see figure15).

```
void test_memory_range(struct memory_range_s *range) {
    XStatus status;

    /* This application uses print statements instead of xil_printf/printf
     * to reduce the text size.
     *
     * The default linker script generated for this application does not have
     * heap memory allocated. This implies that this program cannot use any
     * routines that allocate memory on heap (printf is one such function).
     * If you'd like to add such functions, then please generate a linker script
     * that does allocate sufficient heap memory.
     */

    print("Testing memory region: "); print(range->name);  print("\n\r");
    print("    Memory Controller: "); print(range->ip);  print("\n\r");
    print("         Base Address: 0x"); putnum(range->base); print("\n\r");
    print("                 Size: 0x"); putnum(range->size); print (" bytes \n\r");

    status = Xil_TestMem32((u32*)range->base, 1024, 0xAAAA5555, XIL_TESTMEM_ALLMEMTESTS);
    print("          32-bit test: "); print(status == XST_SUCCESS? "PASSED!":"FAILED!"); print("\n\r");

    status = Xil_TestMem16((u16*)range->base, 2048, 0xAA55, XIL_TESTMEM_ALLMEMTESTS);
    print("          16-bit test: "); print(status == XST_SUCCESS? "PASSED!":"FAILED!"); print("\n\r");

    status = Xil_TestMem8((u8*)range->base, 4096, 0xA5, XIL_TESTMEM_ALLMEMTESTS);
    print("           8-bit test: "); print(status == XST_SUCCESS? "PASSED!":"FAILED!"); print("\n\r");
```

**Fig 13: Memory testing application analysis on SDK platform**

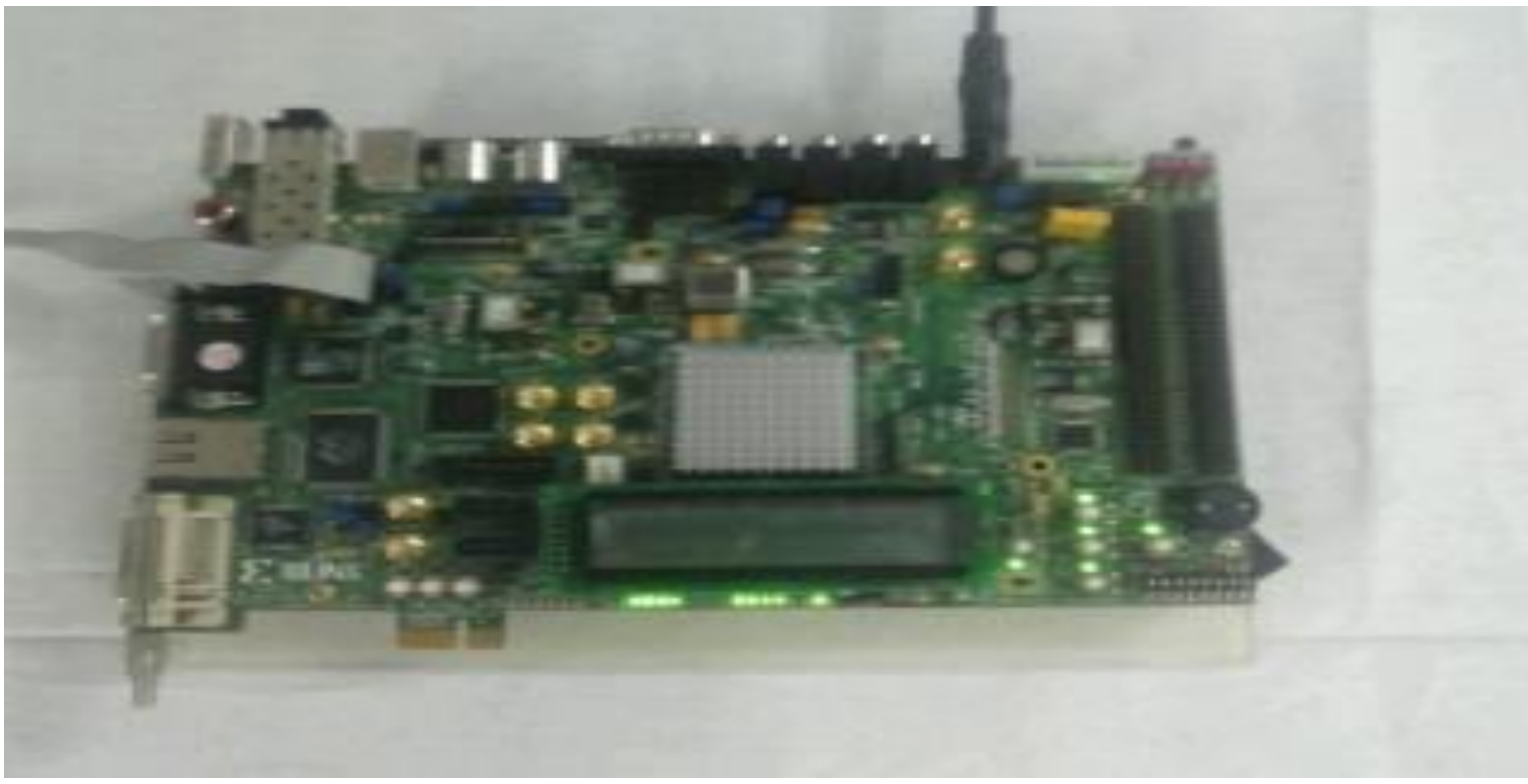

**Fig 14: Xup-5 FPGA Board**

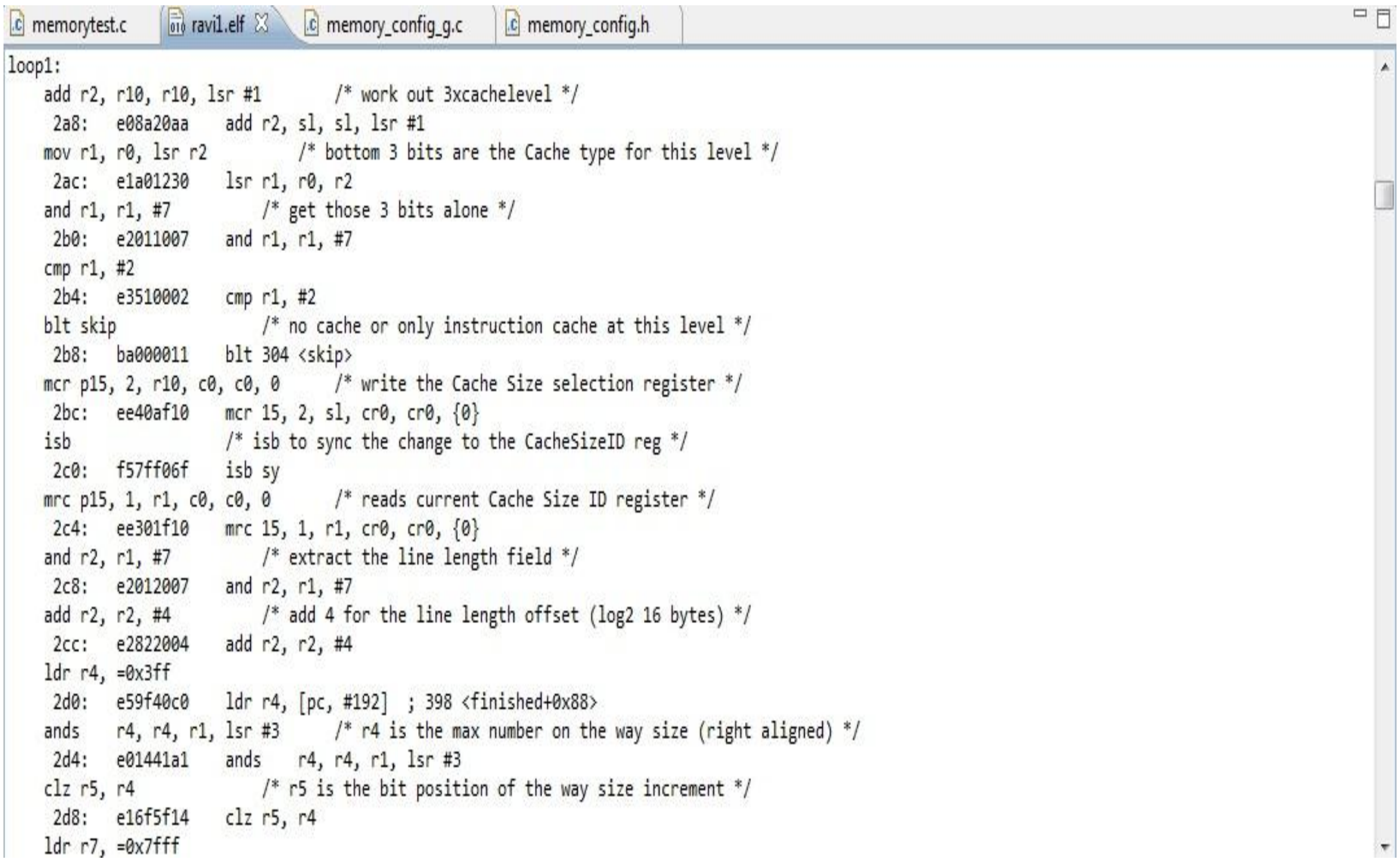

```
memorytest.c   ravi1.elf   memory_config_g.c   memory_config.h
loop1:
    add r2, r10, r10, lsr #1        /* work out 3xcachelevel */
     2a8:   e08a20aa    add r2, sl, sl, lsr #1
    mov r1, r0, lsr r2          /* bottom 3 bits are the Cache type for this level */
     2ac:   e1a01230    lsr r1, r0, r2
    and r1, r1, #7          /* get those 3 bits alone */
     2b0:   e2011007    and r1, r1, #7
    cmp r1, #2
     2b4:   e3510002    cmp r1, #2
    blt skip                /* no cache or only instruction cache at this level */
     2b8:   ba000011    blt 304 <skip>
    mcr p15, 2, r10, c0, c0, 0      /* write the Cache Size selection register */
     2bc:   ee40af10    mcr 15, 2, sl, cr0, cr0, {0}
    isb                 /* isb to sync the change to the CacheSizeID reg */
     2c0:   f57ff06f    isb sy
    mrc p15, 1, r1, c0, c0, 0       /* reads current Cache Size ID register */
     2c4:   ee301f10    mrc 15, 1, r1, cr0, cr0, {0}
    and r2, r1, #7          /* extract the line length field */
     2c8:   e2012007    and r2, r1, #7
    add r2, r2, #4          /* add 4 for the line length offset (log2 16 bytes) */
     2cc:   e2822004    add r2, r2, #4
    ldr r4, =0x3ff
     2d0:   e59f40c0    ldr r4, [pc, #192]  ; 398 <finished+0x88>
    ands    r4, r4, r1, lsr #3      /* r4 is the max number on the way size (right aligned) */
     2d4:   e01441a1    ands    r4, r4, r1, lsr #3
    clz r5, r4              /* r5 is the bit position of the way size increment */
     2d8:   e16f5f14    clz r5, r4
    ldr r7, =0x7fff
```

**Fig 15: ISA Simulation result of specific application**

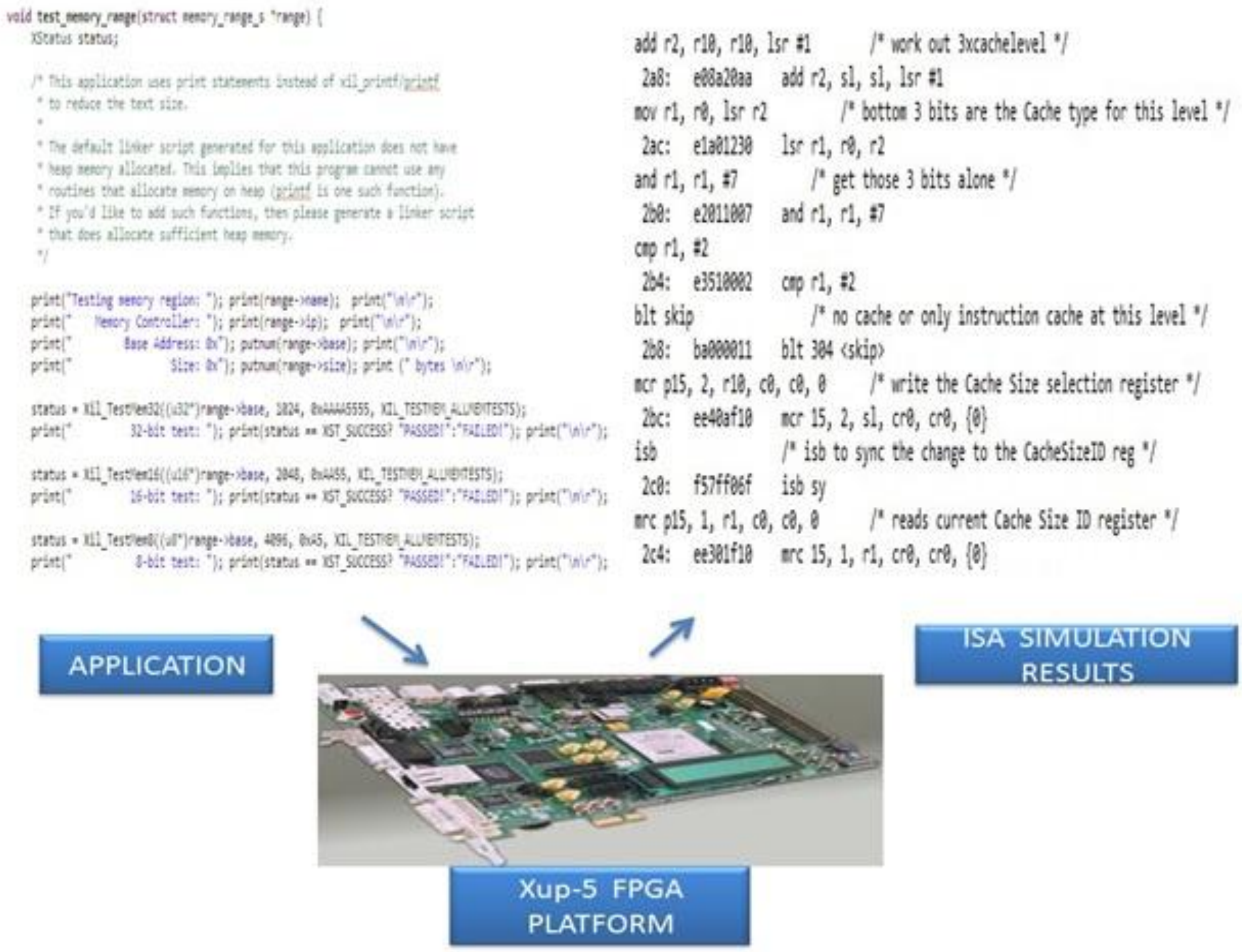

**Fig 16: Application specific simulation analysis**

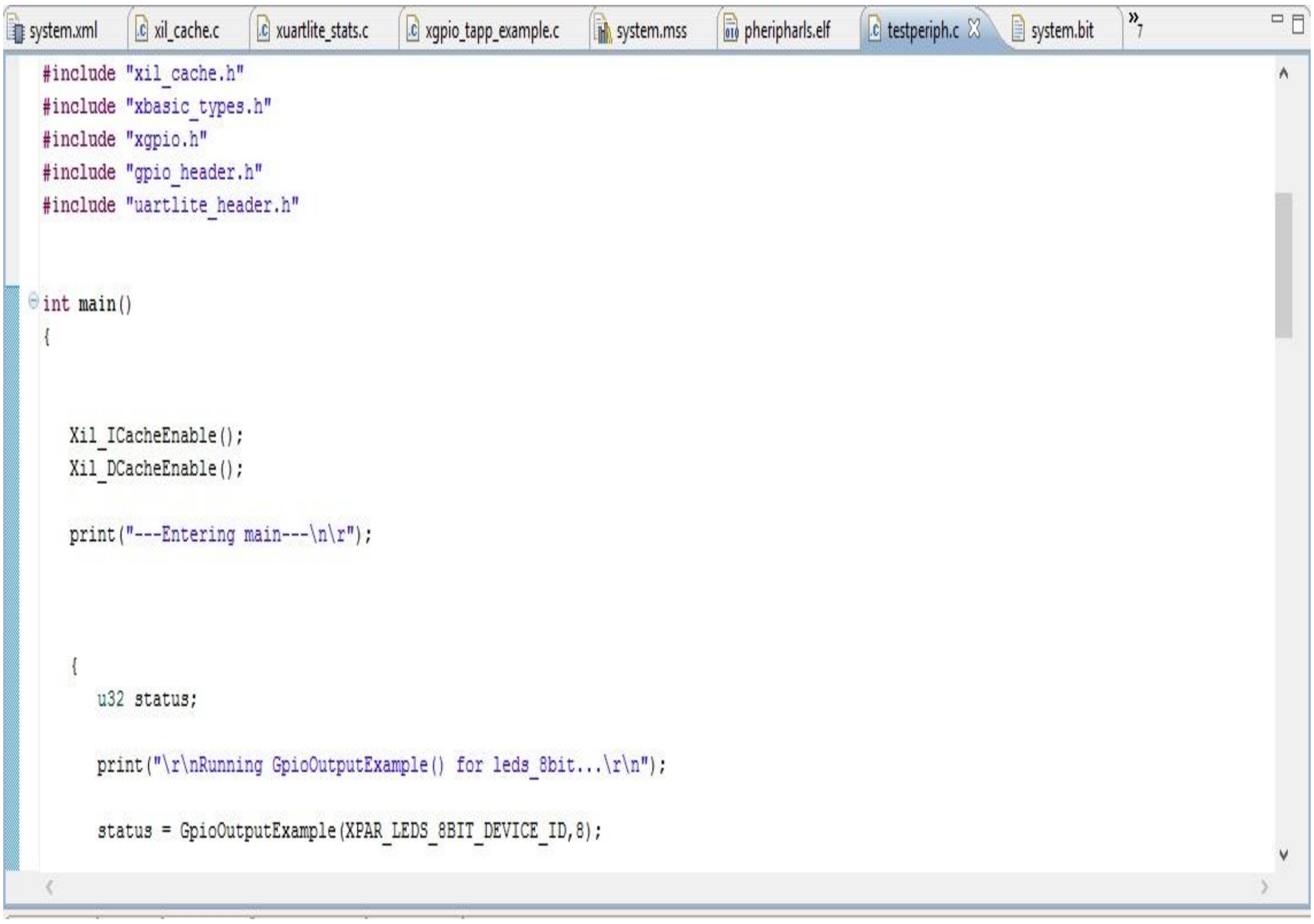

```
#include "xil_cache.h"
#include "xbasic_types.h"
#include "xgpio.h"
#include "gpio_header.h"
#include "uartlite_header.h"


int main()
{


   Xil_ICacheEnable();
   Xil_DCacheEnable();

   print("---Entering main---\n\r");



   {
      u32 status;

      print("\r\nRunning GpioOutputExample() for leds_8bit...\r\n");

      status = GpioOutputExample(XPAR_LEDS_8BIT_DEVICE_ID,8);
```

**Fig 17: Peripheral test application**

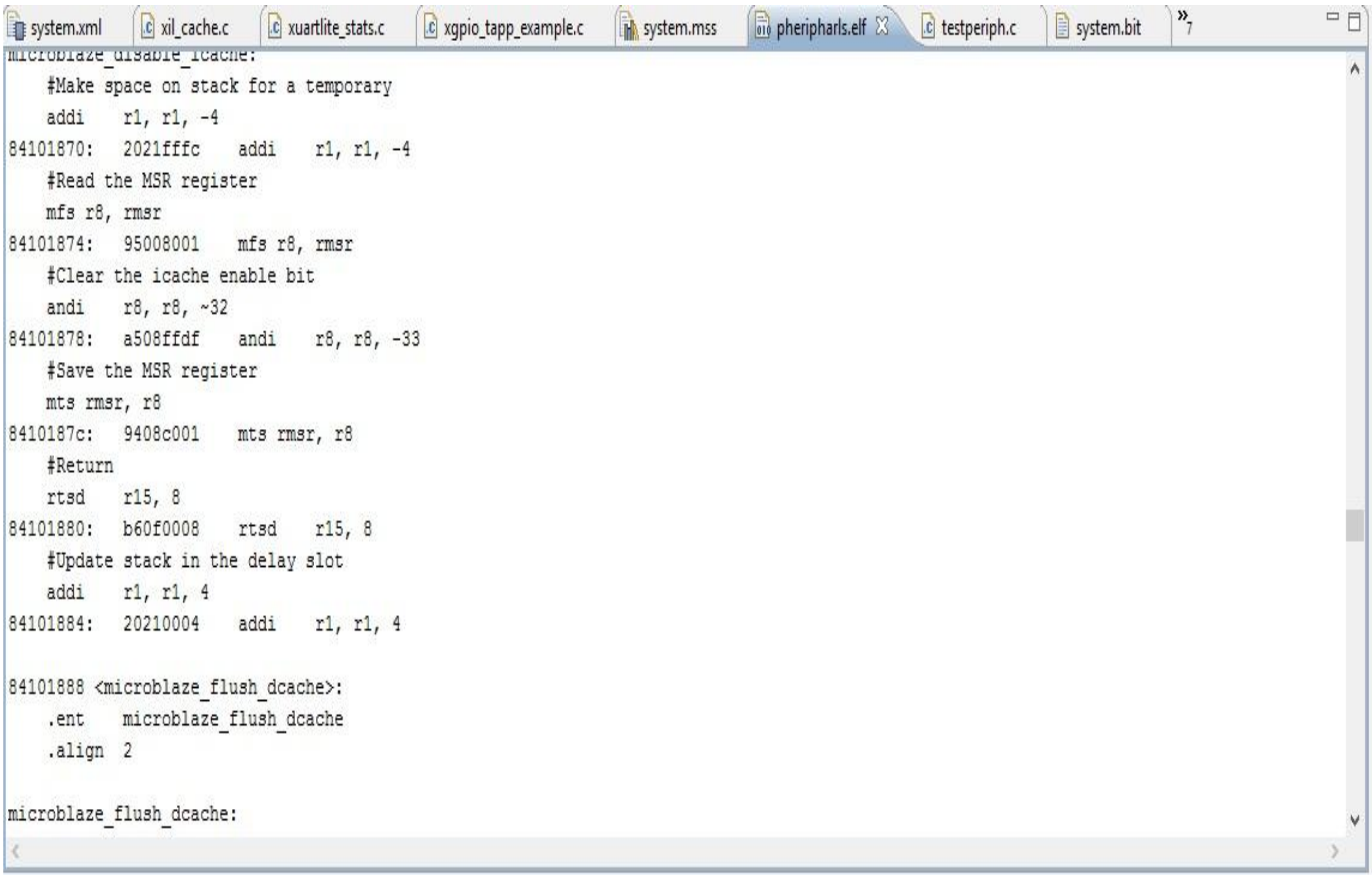

```
microblaze_disable_icache:
    #Make space on stack for a temporary
    addi    r1, r1, -4
84101870:   2021fffc    addi    r1, r1, -4
    #Read the MSR register
    mfs r8, rmsr
84101874:   95008001    mfs r8, rmsr
    #Clear the icache enable bit
    andi    r8, r8, ~32
84101878:   a508ffdf    andi    r8, r8, -33
    #Save the MSR register
    mts rmsr, r8
8410187c:   9408c001    mts rmsr, r8
    #Return
    rtsd    r15, 8
84101880:   b60f0008    rtsd    r15, 8
    #Update stack in the delay slot
    addi    r1, r1, 4
84101884:   20210004    addi    r1, r1, 4

84101888 <microblaze_flush_dcache>:
    .ent    microblaze_flush_dcache
    .align  2

microblaze_flush_dcache:
```

**Fig 18: Peripheral appplication Simulation results**

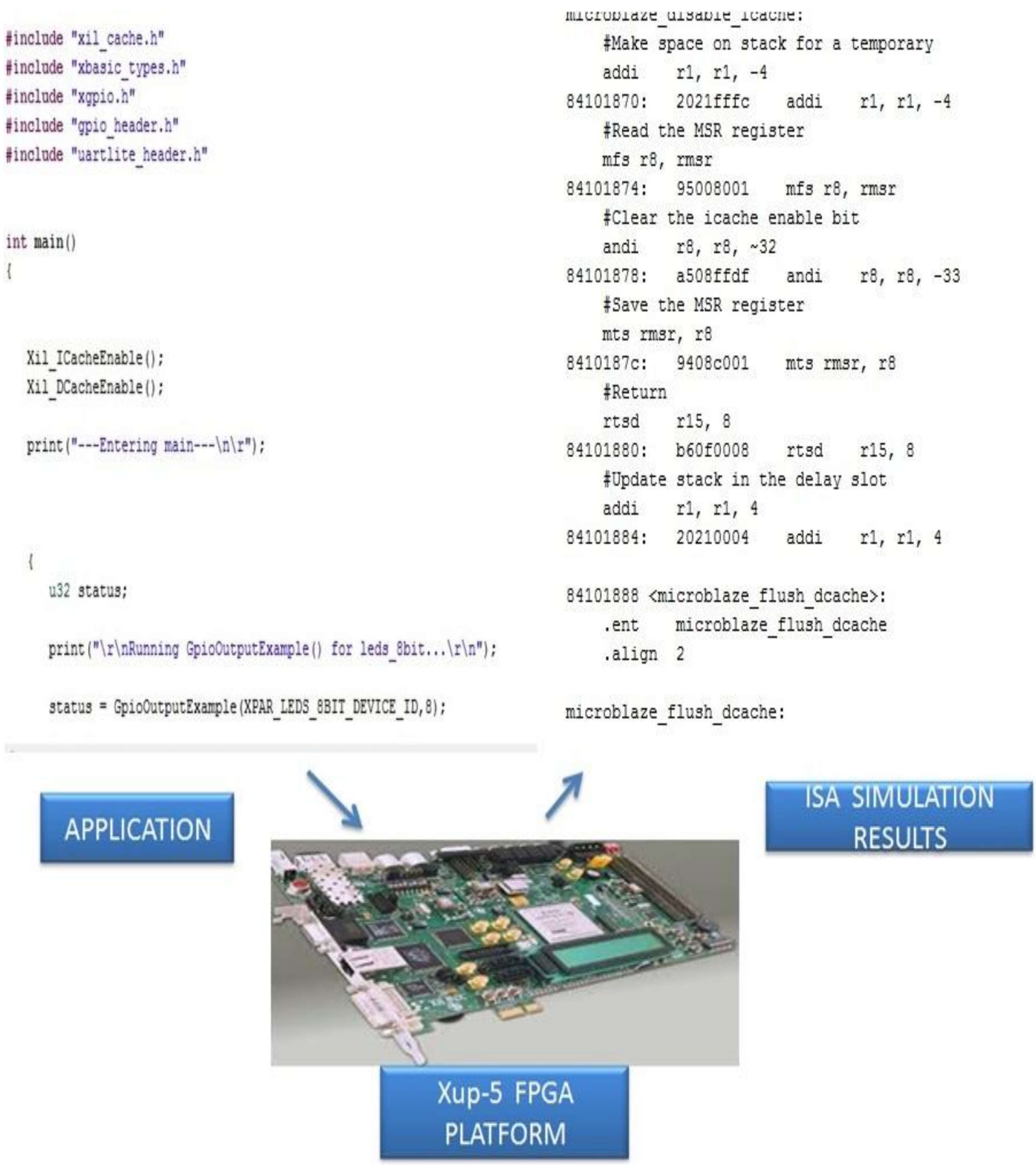

**Fig 19: Peripheral application simulation analysis with xup-5 FPGA BOARD**

Memory testing application (see figure 13) used on specific target hardware (see figure 14). SDK examined the hardware specification file and tested the memory application with xup-5 FPGA board. It generates the .elf file and elf contains the ISA behavior for specific application. After the simulation process (see figure 16) we get efficient ISA result (see figure 15). Peripherals testing another application (see figure 17) is used for simulation. It generates the simulation results (see figure 18). The simulation process is to analyze on xup-5 fpga device with custom hardware environment (see figure19).

## 5. CONCLUSION

Our main focus in this paper is application specific simulation analysis with specific hardware environment. Hardware resources like processor; memories etc. easily implemented and analyzed custom instruction set behavior for specific application. Various standard applications analyzed on xup-5 FPGA board with specific hardware environments. The ISA behaviour is also analyzed. After this simulation process application specific results are used for high performance embedded system.

## 6. REFERENCES

[1] Kucukcakar, K. An ASIP design methodology for embedded systems. In proc of: Hardware/Software Co-design, 1999.

[2] Jain, M.K., Balakrishnan, M., Kumar, A., "ASIP Design Methodologies: Survey and Issues", VLSI, 14th

International Conference on VLSI Design (VLSID '01), 2001.

[3] Hartmann, M., Raghavan, P., Perre, V. D., L., Agrawal, P., Memristor-Based (ReRAM) Data Memory Architecture in ASIP Design, IEEE, Digital System Design (DSD), 2013,Euromicro Conference. pp.795 – 798.

[4] Sharma, A., Sutar, S., Sharma, V.K., Mahapatra K.K. An ASIP for image enhancement applications in spatial domain using LISA,2011, pp.175-179.

[5] Fathy A., Isshiki T.,Li D., Kunieda H. Custom Instruction Synthesis Framework for Application Specific Instruction-Set Processor with HW, IC-ICTES in Ayutthaya, 2014.

[6] Xilinx tool Available from http//www.xilinx.com.

[7] Qiu, J., Gao, X., Jiang, Y., Xiao, X. An ultra-fast hybrid simulation framework for ASIP, Electronics, Circuits and Systems (ICECS), 2011 18th IEEE International Conference. pp.711 – 714.

[8] Hassan**,** H. M., Mohammed, K. and Shalash**,** A. F. Implementation of a reconfigurable ASIP for high throughput low power DFT/DCT/FIR engine, engine EURASIP Journal on Embedded Systems, 2012.